%% file: main.tex
\documentclass{llncs}
\usepackage[T1]{fontenc}

\AtBeginDocument{%
  }

\usepackage[svgnames]{xcolor}

\makeatother
\usepackage[misc]{ifsym} 
\usepackage{amsmath}
\usepackage{amssymb}
\usepackage{amsfonts}
\usepackage{graphicx}
\usepackage{textcomp}
\usepackage{tikz}
\usepackage{filecontents}
\usepackage{color, colortbl}
\usepackage[ruled,french,boxed,vlined,linesnumbered]{algorithm2e}
\usepackage[noend]{algpseudocode}
\usepackage{siunitx}
\usepackage{balance}
\usepackage{makecell}
\usepackage[skip=4pt]{caption}
\usepackage{diagbox}
\usepackage{float}
\usepackage[colorlinks=true, linkcolor=Maroon, urlcolor=Maroon]{hyperref}
\usepackage{threeparttable}
\usepackage{url}
\usepackage{verbatim} 
\usepackage{xspace}
\usepackage{totpages}
\usepackage{varwidth}
\usepackage{subfigure}
\usepackage{makecell}
\usepackage{multirow}
\usepackage{flushend}                    
\usepackage[shortlabels]{enumitem}
\usepackage[noend]{algpseudocode}
\usepackage{cleveref}

\usepackage{booktabs} % 用于更美观的表格线

\usepackage{subcaption} 
\usepackage{pythonhighlight}

\newcommand{\sysname}[0]{\texttt{FlexPie}\xspace}

\newcommand{\hwsay}[1]{} %{\cblue{[hw: #1]}}

\newenvironment{parafont}{\fontfamily{ptm}\selectfont}{}
\newcommand{\Para}[1]{\vspace{2pt}\noindent\begin{parafont}\textbf{\textit{#1}}\end{parafont}}

\usepackage{ragged2e}

\begin{document}

\title{\sysname{}: Accelerate Distributed Inference on Edge Devices with Flexible Combinatorial Optimization}
\author{}\institute{}
\author{Runhua Zhang$^*$,
Hongxu Jiang\Letter$^*$,
Jinkun Geng$^+$,
Yuhang Ma$^*$,
Chenhui Zhu$^*$,
Haojie Wang$^\dagger$,\\
$^*$Beihang University, $^+$Stanford University,$^\dagger$Tsinghua University\\
\{rhzhang20, jianghx\}@buaa.edu.cn, gjk1994@stanford.edu,\\ 
\{buaa\_mayuhang, zhuch0401\}@buaa.edu.cn, wanghaojie@tsinghua.edu.cn
}

\maketitle            

\sloppypar

\input{abstract}

\input{intro}

\input{background}

\input{design}

\input{experimental}

\input{related}

\input{conclusion}

\bibliographystyle{plain}

\bibliography{ref}

\end{document}

%% file: abstract.tex
\begin{abstract}

%The rapid development of deep learning leads to the emergence of many novel IoT (Internet of Thing) applications. These applications usually deploy the pre-trained DNN (Deep Neural Network) models to multiple (usually $4\sim6$) edge devices to jointly execute the inference tasks. However, the conventional deployment tends to yield undesirable inference time. We have studied the problem and identified that the reason is mainly due to the model partition strategy adopted by the inference engine. Prior inference frameworks either adopt a fixed model partition scheme (e.g., One-dim partition or 2D-grid partition), or integrate limited optimization (e.g., layerwise optimization or simple layer fusion), which fails to achieve the optimal performance towards the different model layers or testbed settings. 

The rapid advancement of deep learning has catalyzed the development of novel IoT applications, which often deploy pre-trained deep neural network (DNN) models across multiple edge devices (typically 4$\sim$6) for collaborative inference. However, conventional deployment methods frequently result in suboptimal inference times. Our study identifies that the inefficiency primarily stems from the model partitioning strategy employed by inference engines. Previous frameworks often rely on fixed partitioning schemes (e.g., one-dimensional or 2D-grid partitions) or limited optimizations (e.g., layer-wise adjustments or simple layer fusion), which fail to deliver optimal performance across varying model layers and testbed configurations. In this paper, we propose \sysname, a solution to accelerate distributed inference on edge devices through \emph{flexible combinatorial optimization}. \sysname integrates an automated optimization procedure based on a data-driven cost model and dynamic programming, which efficiently finds an optimized model partition scheme in huge combinatorial spaces. Our evaluation on four commonly used DNN benchmarks demonstrates that \sysname reduces inference time, achieving up to a 2.39× speedup over state-of-the-art methods.

\end{abstract}

\keywords{Edge Device, Distributed Inference, Model Partition, Combinatorial Optimization, Data-Driven, Dynamic Programming} 

%% file: intro.tex
\section{Introduction}
The past decade has witnessed a rapid development of deep learning, leading to a variety of novel applications in IoT (Internet of Things) scenarios, including image processing~\cite{abdel2020image,hu2017iot}, video analysis~\cite{long2017edge}, wearables~\cite{bi2019ramt} and so on. Due to network congestion and long physical distances, cloud services cannot meet low latency requirements~\cite{liu2023joint}. Therefore, typical IoT applications tend to train the ML model in beefy clusters (e.g., cloud VMs or bare-metal machines), but the inference computation is usually offloaded to the edge device for the sake of real-time responsiveness~\cite{zhou2019adaptive,chen2024edgeci} and data privacy~\cite{dhar2022studying}. 

While edge-based inference offers more timely results than its cloud-based counterpart, it also leads to unique challenges. Because of the growing size of DNN models, deploying a single-edge device to load and process an entire model has become impractical. Consequently, distributed inference is required to enable collaborative execution of inference tasks. Today's edge applications~\cite{zhang2021deepslicing,zhao2018deepthings,mohammed2020distributed,dey2019embedded} usually employ 4$\sim$6 nodes\footnote{Due to the practical constraints such as energy consumption, today's edge applications usually conduct small-scale (4$\sim$6 nodes) distributed inference, which is different from the cloud-based distributed computing that employs 10s--100s nodes.} to jointly execute the inference for efficiency. Commonly, DNN feature maps are partitioned into different edge devices. Each edge device hosts a partial model and exchanges necessary parameters with others throughout the layer-based inference process.

Reviewing the existing distributed inference solutions, we can divide them into two categories. One line of prior works~\cite{zhao2018deepthings,zhou2019adaptive,mao2017modnn} adopt a fixed partition scheme, regardless of model shapes and testbed settings (e.g., the number of edge devices to run the inference). For instance, DeepSlicing~\cite{zhang2021deepslicing} and MoDNN~\cite{mao2017modnn} adopt the One-dim partition scheme (i.e., partition the model by InW or InH or OutC dimension) whereas DeepThings~\cite{zhao2018deepthings} uses the 2D-grid partition scheme to run the distributed inference task. However, we consider the flexible model partition as a necessity due to two observations in our measurement study (details in~\S\ref{sec:micro}): (1) When running inference computation on the same group of edge devices, different DNN layers yield their optimal inference time with different model partition schemes; (2) when running the inference computation for the same layer of DNN, the varying number of edge devices leads to different optimal model partition schemes, i.e., the optimal partition scheme obtained from one testbed will no longer be the optimal after we switch to another testbed. 

The other line of works~\cite{chen2024edgeci,dey2019embedded,mohammed2020distributed,zhou2019adaptive} introduce limited flexibility into the inference framework. For instance, DINA~\cite{mohammed2020distributed} and PartialDI~\cite{dey2019embedded} adopt layerwise optimization and allow the framework to use different partition schemes for every model layer. However, it ignores the inter-layer dependency, which can be further optimized with layer fusion. On the other hand, EdgeCI~\cite{chen2024edgeci} and AOFL~\cite{zhou2019adaptive} explore the opportunities for layer fusion, but its fusion optimization is only applicable for a single partition scheme. While combining both layerwise optimization and fusion-based optimization seems straightforward, such a combination usually leads to a very large search space in practical deployment. The simple exhaustive search can be time-consuming and requires much expertise, which we believe is the main reason that discourages previous work from combining both \emph{flexible layerwise partition} and \emph{opportunistic layer fusion}.

\Para{Our goal.} We develop \sysname, which targets the scenarios of edge-based inference that employ multiple (4$\sim$6) devices to jointly compute the inference results. Such scenarios have become prevalent in many practical applications~\cite{4orin,2orin}. \sysname fully considers the optimization opportunities of layerwise optimization and inter-layer fusion to generate the desirable model partition scheme. In order to efficiently find the optimal scheme in a large design space, \sysname incorporates an automatic model partition strategy, named \emph{flexible combinatorial optimization} (FCO). In general, FCO runs the dynamic programming process and data-driven cost model to rapidly pick out the best model partition scheme to deploy for a given model and testbed. To implement FCO, \sysname consists of two main components. 

\Para{Data-Driven Cost estimator (CE, \S\ref{sec:gdbt-estimator}).} \sysname employs Gradient Boosting Decision Tree (GBDT) models to estimate the inference time cost for the model layers. We implement the GBDT based on XGBoost~\cite{chen2016xgboost} and collect data traces for edge-based inference under various settings, and then we use these data traces to train two GBDT models (estimators), namely i-Estimator and s-Estimator. The estimators take as input the DNN layer's metadata (including the height, width, and number of input/output channels, etc.) and the testbed information (including the bandwidth, communication topology, etc.) and then output an estimated time cost. The i-estimator estimates the time cost to complete the inference computation on this model layer; the s-estimator estimates the time cost for all the nodes in the testbed to complete the synchronization of the model layer. Estimated time costs provide useful guide information for \sysname to make choices among partition schemes to minimize the overall time cost (we design a dynamic programming algorithm to achieve this).

\Para{Dynamic partition planner (DPP, \S\ref{sec:dpp}).} Driven by the i-Estimators and s-Estimator, we continue to design and implement a dynamic programming (DP) algorithm to decide the partition scheme for a given model and testbed. Taking the typical DNN model as an example, the DPP starts the DP process from the last layer of the DNN model. By contacting the i-Estimator, DPP knows the estimated inference time cost when it decides to adopt a specific partition scheme (e.g., InW-based, InH-based, OutC-based, 2D-grid, etc.) for one model layer of some specific shape. By contacting the s-Estimator, DPP knows the estimated cost to synchronize one model layer of some specific shape among the nodes of the cluster. Based on the i-Estimator and the s-Estimator, DPP can search and evaluate a series of partition schemes in the DP search space.  DPP will keep the promising candidate partitions which can lead to the lowest overall inference time and keep traversing forwards, and finally output a desirable partition scheme. Besides, during the DP process, we have also incorporated the pruning strategy to reduce the search space and find the optimal partition scheme (i.e., the scheme yielding the lowest overall inference time) more efficiently.

\Para{Evaluation.} We compare \sysname with five partition schemes, i.e., InW/InH-based, OutC-based, 2D-grid, layerwise and fused-layer partition. We conduct inference with four models under different testbed settings. We find that, (1) The 2D-grid partition outperforms One-dim partition (InH/InW-based and OutC-based) when we conduct the distributed inference on the 4-node testbed. However, the InH/InW-based partition outperforms the 2D-grid and OutC-based partition when we switch to the 3-node testbed. Running on different testbeds, these typical partition schemes can all lead to an imbalance of model partition, and thus are not generally applicable to performance improvement. (2) Both layerwise optimization and data fusion can gain better inference speedup than the fixed partition scheme. However, simply using either strategy fails to yield optimal performance. By comparison, \sysname combines both advantages and can adaptively choose the optimal partition scheme for different models under different testbed settings. Therefore, \sysname outperforms all the baselines across multiple benchmarks with a speedup of 1.10-2.39$\times$.

%% file: background.tex
\section {Background and Motivation}
\label{sec-back}
\subsection{Distributed Inference and Model Partition}

\begin{figure}[!t]
    \centering
    \begin{minipage}{0.48\textwidth}
        \centering
        \includegraphics[width=\linewidth]{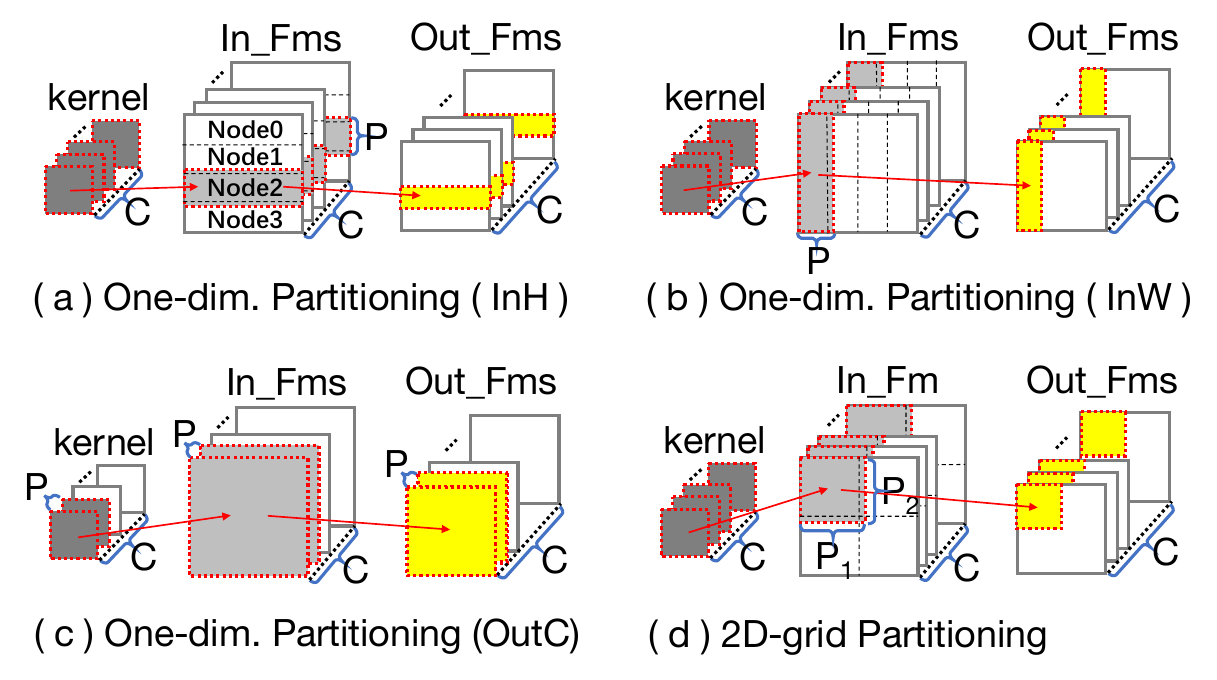}
        \caption{Example of parallelizing depthwise separable convolution}
        \label{fig:model-partition}
    \end{minipage}\hfill    
    \begin{minipage}{0.47\textwidth}
        \centering
        \includegraphics[width=\linewidth]{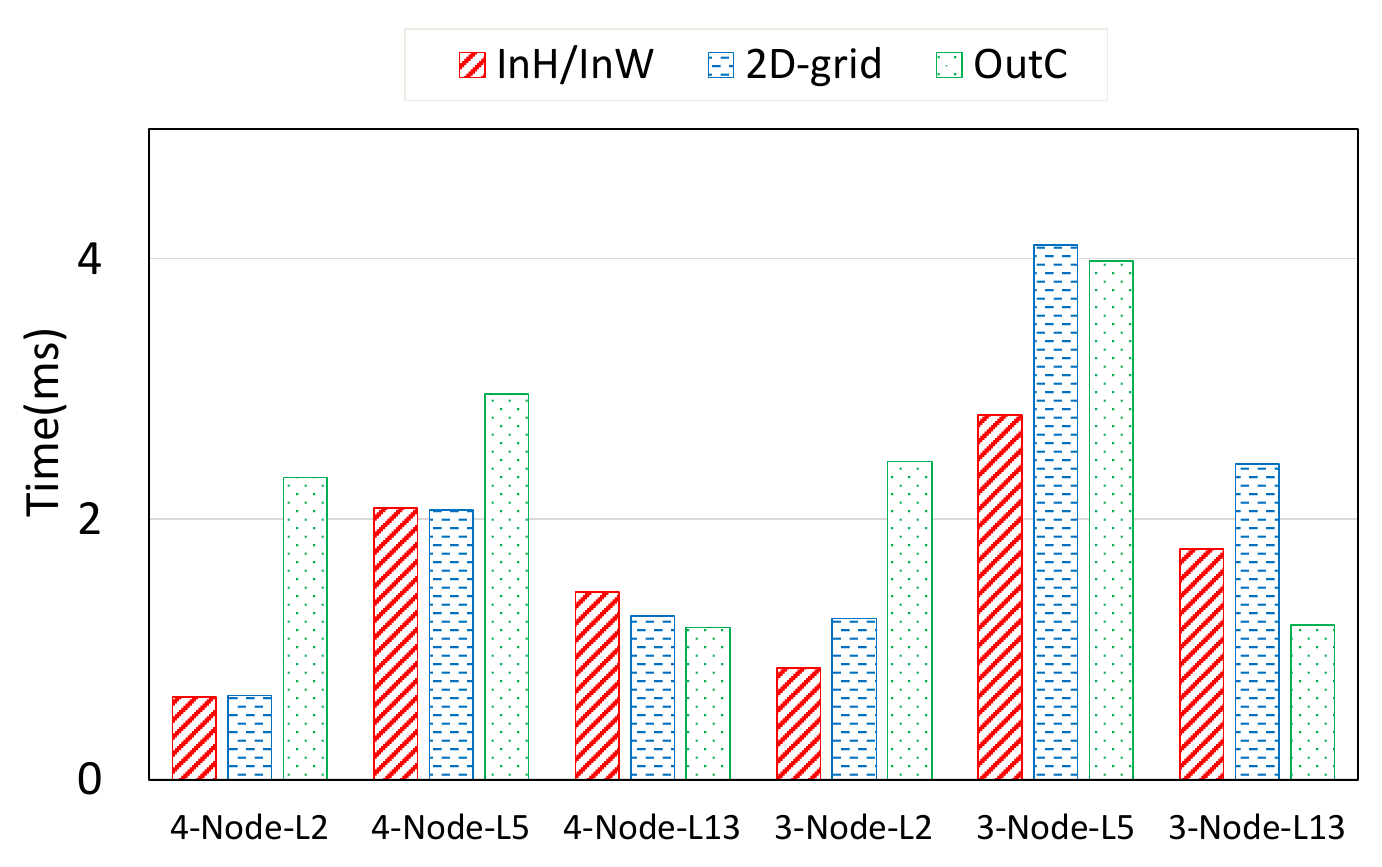}
        \caption{Micro-bench test}
        \label{fig:micro-bench}
    \end{minipage}
\end{figure}

% \begin{figure}[!t]
%     \centering
%     \begin{minipage}{0.45\textwidth}
%         \centering
%         \includegraphics[width=\linewidth]{fig/micro-bench.pdf}
%         \caption{Micro-bench test}
%         \label{fig:micro-bench}
%     \end{minipage}\hfill
%     \begin{minipage}{0.55\textwidth}
%         \centering
%         \includegraphics[width=\linewidth]{fig/FCO.pdf}
%         \caption{Illustration of \emph{combinatorial flexible partition} scheme}
%         \label{fig:FCO}
%     \end{minipage}\hfill
%     % \vspace{-0.5cm}
% \end{figure}

%Distributed DNN inference becomes increasingly necessary in edge-based applications because the inference model is growing in size but the applications have stringent real-time latency requirement. By employing multiple edge devices to jointly execute the inference task, the completion time is expected to be reduced to achieve the desirable application responsiveness. 

% While conducting distributed inference, a straightforward question that arises is to decide the model partition scheme across multiple edge devices. 

Figure~\ref{fig:model-partition} illustrates the commonly used partition schemes in practice. Typically in a DNN model, there are two main categories of partition schemes, namely One-dim partition (Figure~\ref{fig:model-partition}(a)-\ref{fig:model-partition}(c)) and 2D-grid partition(Figure~\ref{fig:model-partition}(d)). In One-dim partition, there are three commonly-used sub-categories, i.e., InH-based partition, InW-based partition and OutC-based partition, which indicate that the partition is conducted according to the height of the input feature map (InH), the width of the input feature map (InW), and the channel of the output feature map (OutC). As for the 2D-grid partition, it partitions the model on both the InH dimension and the InW dimension for the sake of load balance. Notably, the InC-based partition is not typically used to partition the model, because it introduces costly gather operations and leads to undesirable performance.

\subsection{Partition Scheme Affects Performance}

\label{sec:micro}
While One-dim and 2D-grid partition schemes are both generally applicable to inference models, we find that they lead to distinctly different completion times when we conduct the inference with different layers under different testbed settings. Figure~\ref{fig:micro-bench} demonstrates this using a series of micro-bench tests. We conduct inference on MobileNet with different partition schemes and measure the completion inference time on different layers. In the first 3 groups of tests, i.e., 4-Node-L2, 4-Node-L5, 4-Node-L13, we run the inference with 4 edge devices, connected with SRIO at a bandwidth of 5Gb/s. The three groups of tests conduct inference with different layers, i.e., the 2nd layer (L2), the 5th layer (L5) and the 13th layer (L13). Obviously, different layers yield their optimal time with different partition schemes. L2 and L5 prefer the InH/InW-based partition and the 2D-grid partition, whereas L13 prefer the OutC-based partition. Next, we change the testbed settings from using 4 nodes to using 3 nodes, then we get the other three groups of tests, i.e., 3-Node-L2, 3-Node-L5 and 3-Node-L13. Comparing with the first 3 groups we can see that, even towards the same layer, the optimal partition schemes vary under different testbed settings. For example, L5 can yield an optimal inference time with 2D-grid partition in the 4-node setting, but the 2D-grid partition gives the worst inference time for L5 under the 3-node setting. 

From the micro-bench tests, we realize that there are no \emph{one-size-fits-all} partition schemes. Even in the same model, the different layers may prefer different partition schemes. Besides, the optimality of the partition schemes can also vary across different testbed settings. Therefore, we are motivated to design flexible partition schemes for different model layers under various testbed settings.

\subsection{Trade-off between Computation and Communication }
\label{sec:trade-off}

% \begin{figure}[t]
% 	\centering
% 	\subfigure[One-dim.(InH)]{
% 		\label{fig:one-dim-trade-off} 
% 	\includegraphics[width=0.2\textwidth]{./fig/one dim_trade-off.pdf}
% 	}
%     \subfigure[2D-grid]{
% 	\label{fig:2d-grid-trade-off} 
%         \includegraphics[width=0.2\textwidth]{./fig/2d-grid_trade-off.pdf}
% 	}
% 	\caption{Trade-off between (redundant) computation and communication}
% 	\label{fig:trade-off} 
%         % \vspace{-0.5cm}
% \end{figure}

One-dim and 2D-grid partitions lead to different overlaps between computation (on each edge device) and communication (across multiple edge devices). Figure~\ref{fig:model-partition}(a) illustrates the One-dim partition scheme (InH-based) when we conduct distributed inference with 4 nodes (Node0-Node3). The inference engine first uses the kernel and input feature maps (In\_Fms) to inference, computing output feature maps (Out\_Fms). Then the Out\_Fms become the In\_Fms for the next layer. From Figure~\ref{fig:model-partition}(a) we can see that Node2 still requires some boundary data to be transferred from other nodes (e.g. Node1 and Node3) to perform a new layer inference. 

% JSA
\iffalse 
\begin{table*}[!t]
\renewcommand\arraystretch{1}
	\centering
	\caption{Comparison with prior frameworks}
	\label{frameworks}
\footnotesize
	\begin{tabular}{ccccc}
		\hline
		&Layer-Wise &  \makecell{Fused-Layer}  & Combinatorial Optimization & Data-Driven \\
		\hline
            MoDNN~\cite{mao2017modnn}                 &  $\times$  (One-dim.) & $\times$    &  $\times$   &  $\times$       \\
            DeepThings~\cite{zhao2018deepthings}      &  $\times$ (2D-grid)   & $\times$    &  $\times$   &  $\times$       \\
            DeepSlicing~\cite{zhang2021deepslicing}   & $\times$ (One-dim.)   & $\times$    &  $\times$   &  $\times$       \\
            AOFL~\cite{zhou2019adaptive}              & $\times$ (2D-grid)    & \checkmark  &  $\times$   &  \checkmark     \\
            EdgeCI~\cite{chen2024edgeci}              & $\times$  (One-dim.)  & \checkmark  &  $\times$   &  $\times$       \\
            DINA~\cite{mohammed2020distributed}       & \checkmark            & $\times$    &  $\times$   &  $\times$       \\
            PartialDI~\cite{dey2019embedded}          & \checkmark            & $\times$    &  $\times$   &  $\times$       \\
            \textbf{\sysname}                         &\checkmark             & \checkmark  &  \checkmark & \checkmark      \\
            \hline
	\end{tabular} 
\end{table*}
\fi

Instead of fetching the data from other nodes, which incurs more communication overheads, an alternative way is to let each node conduct redundant inference computation in the previous layer. In Figure~\ref{fig:model-partition}(a), when Node2 is conducting inference, we let it take not only the input from the dashed black rectangle but also the input from the dashed red rectangle. Therefore, Node2 is conducting more computation workload (i.e., Node2 does duplicate computation as Node1 and Node3), but it saves the communication overheads because the required In\_Fms can be obtained from the local node in the next layer. 

The redundant computation may not always be beneficial. For instance, when some model layers have large sizes and the inter-node bandwidth is high, trading computation overheads for communication efficiency does not help accelerate the inference. Therefore, given a specific testbed, \sysname should \emph{flexibly} decide whether or not to conduct redundant computation for each model layer (\S\ref{sec-design}).

Motivated by the existing drawbacks, we develop \sysname to (1) support flexible model partition for each layer, (2) conduct inter-layer optimization to yield better trade-off  between computation and communication, and (3) automate the end-to-end optimization workflow to avoid human tuning effort and efficiently generate the optimal deployment scheme from a huge search space.

%% file: design.tex
\section{\sysname Design}
\label{sec-design}
\subsection{Architecture Overview}
\begin{figure}[t]
	\centering
	\includegraphics[width=0.55\textwidth]{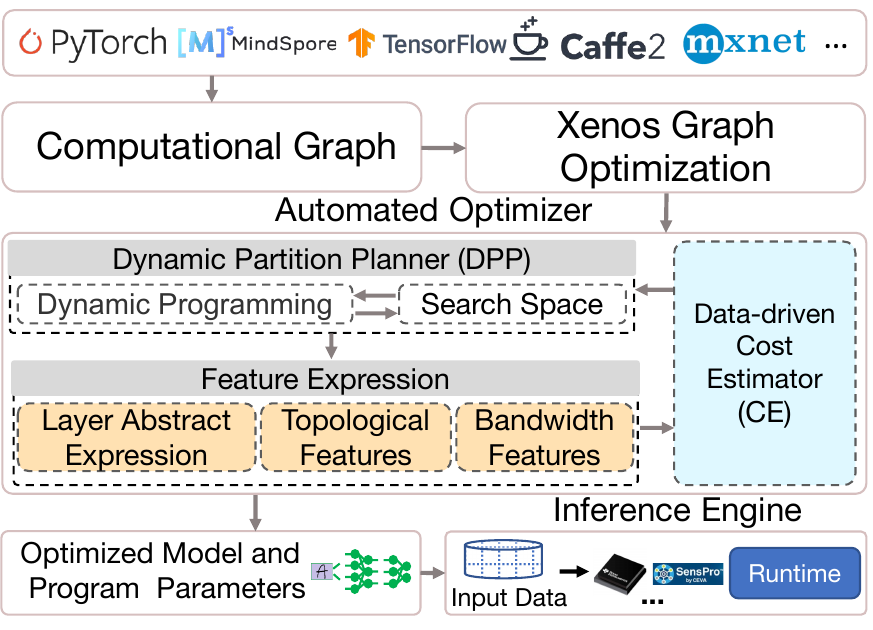}
	\caption{The architecture of \sysname}
	\label{fig:arch}
\end{figure}
Figure~\ref{fig:arch} illustrates the architecture of \sysname. \sysname takes the computation graph as the general intermediate input and can 
support the inference computation for pre-trained models generated from multiple training frameworks (e.g., PyTorch, MindSpore, Tensorflow, etc). Besides, \sysname also integrates pre-optimization strategies from Xenos~\cite{zhang2023xenos} to optimize computation graph before it is fed into the automatic optimizer of \sysname.

The automatic optimizer consists of two major modules, namely the data-driven cost estimator (CE) and the dynamic partition planner (DPP). CE (\S\ref{sec:gdbt-estimator}) is implemented with a Gradient Boosting Decision Tree (GBDT). The GBDT is pre-trained with more than 330K pieces of trace data, collected from running different model inference workloads under a variety of testbed settings. DPP runs a dynamic programming algorithm (\S\ref{sec:dpp}) to search for the efficient model partition scheme. During runtime, DPP contacts CE to get an estimated time cost for the partition scheme in its consideration. Based on the estimated costs, DPP reserves the promising candidate schemes and continues its search. When the DP algorithm is completed, DPP will output the complete model partition scheme with the lowest estimated time cost. Then the inference engine drives multiple edge devices (nodes) to jointly execute the distributed inference computation according to the partition scheme. % We next describe CE and DPP in more detail.

% Specifically, we run different inference models and collect the inference times of these trials. For each trial, we record the shapes of the model (including the sizes of InW, InH, InC, outC, etc) and the testbed settings (including the number of edge nodes, the inter-edge bandwidth, the topology information of the edge cluster, etc). Together with the inference time, we build a piece of sample data corresponding to this trial. 

%From top to bottom, FlexPie first obtains the model calculation graph through ONNX. Secondly, the calculation graph is optimized by Xenos, the previous work, and the optimized calculation graph will be handed over to the Automated Optimizer (AO) to perform model partitioning and optimization. There are three modules in AO, Dynamic Programming Optimizer (DPO), Feature Expression (FE) and Cost Model (CM). DPO performs a combined optimization based on the model Partition Space and the model Equalizer. During the optimization process, the (Feature Expression) PE of the candidate solution will be generated. Finally, the PE will be sent to the Cost model (CM) for prediction, and it will be continuously iterated to generate the partition plan of the entire model. And finally divide the corresponding model parameters.

\subsection{Data-Driven Cost Estimator (CE)}
\label{sec:gdbt-estimator}

% Specifically, we run different inference models and collect the inference times of these trials. For each trial, we record the shapes of the model (including the sizes of InW, InH, InC, outC, etc) and the testbed settings (including the number of edge nodes, the inter-edge bandwidth, the topology information of the edge cluster, etc). Together with the inference time, we build a piece of sample data corresponding to this trial.

To compare different partition schemes and make the choice between them, we need to decide an indicator to measure the cost of each scheme, so that the cost can serve as the guidance for running the dynamic programming algorithm in DPP. Instead of describing the estimated cost explicitly with some formula, which can be complicated when many variables are involved and fail to generalize, we choose a data-driven method and employ a machine learning model to predict the time cost. We use the GBDT as the cost estimator for \sysname due to its simplicity and interpretability~\cite{chen2016xgboost}. 

\begin{figure}[t]
    \centering
    \begin{minipage}{0.53\textwidth}
        \centering
        \includegraphics[width=\linewidth]{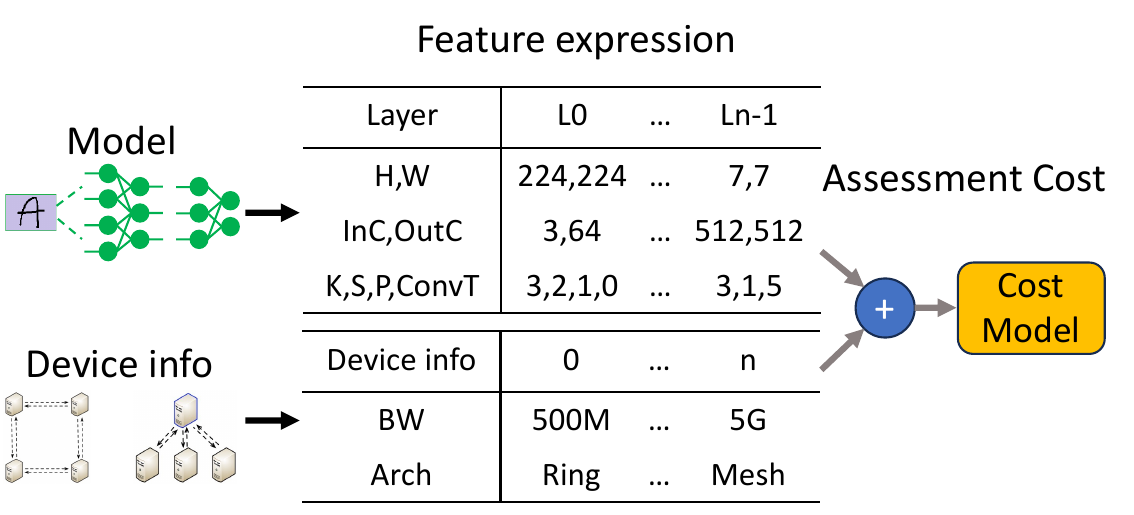}
        \caption{Feature expression}
        \label{fig:feature}
    \end{minipage}\hfill
    \begin{minipage}{0.47\textwidth}
        \centering
        \includegraphics[width=\linewidth]{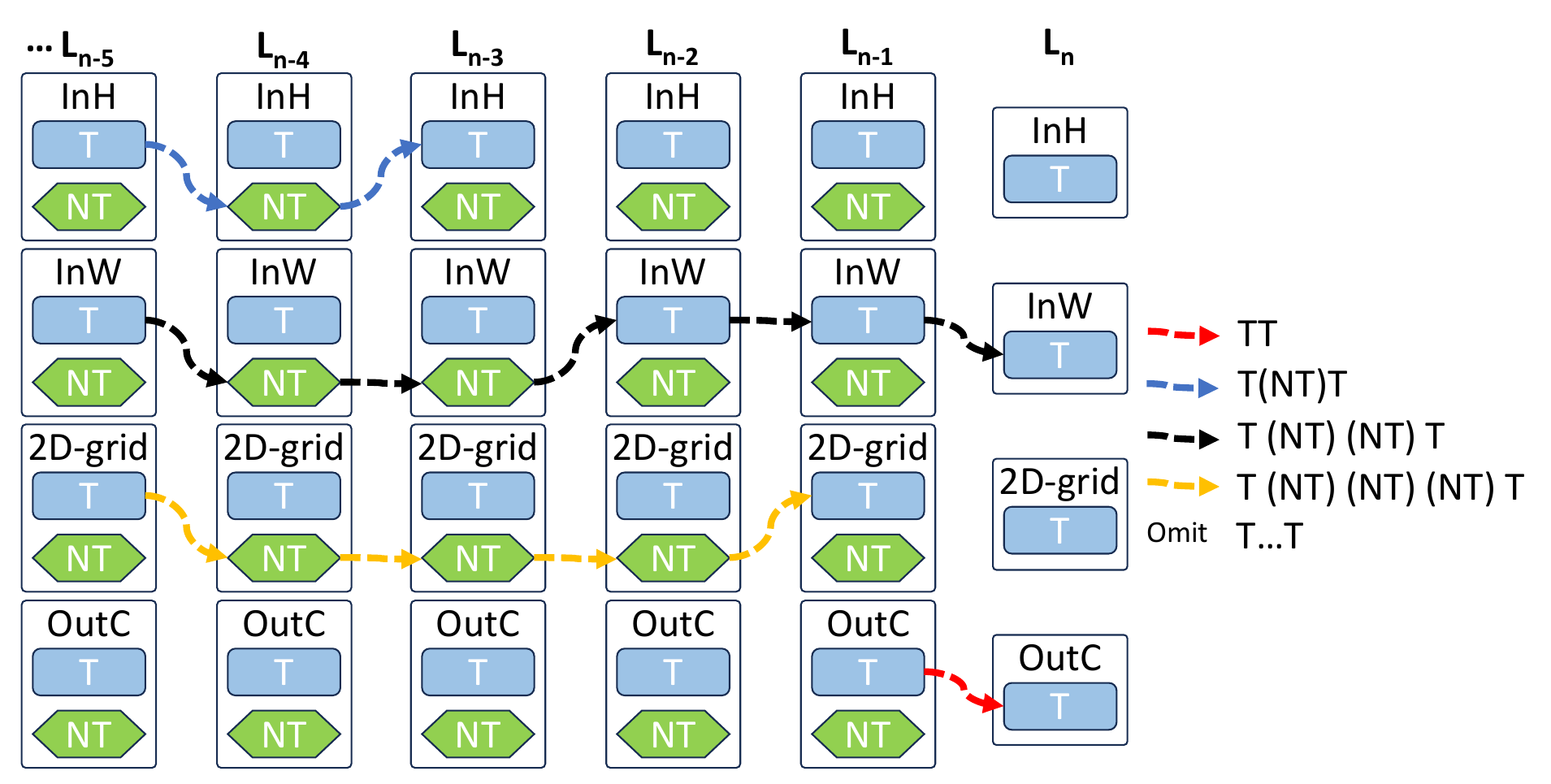}
        \caption{Backtracking process in DPP}
        \label{fig:dpp-illustration}
    \end{minipage}
\end{figure}

The cost estimator takes a group of features as input and outputs an estimated inference time cost to run the distributed inference for a given setting (i.e., running the inference on a specific model layer on a certain testbed). More specifically, we consider three aspects of features (illustrated in Figure~\ref{fig:feature}) to predict the inference time: (1) the shape parameters of the model layer, including InH/OutH, InW/OutW, InC/OutC, K (kernel size), S (stride size), P (padding size),  ConvT (convolution types); (2) the bandwidth between edge devices; (3) the communication architecture adopted by the edge cluster.\footnote{We consider three communication architectures, i.e., ring-based, parameter server (PS)-based, and mesh-based architecture. We have transformed the architecture information into the categorical variable and fed the information to the cost estimator.}

We have run multiple trials to train the i-Estimator and s-Estimator. Each trace data can be represented as a feature vector of 12 dimensions (Figure~\ref{fig:feature}). While training the i-Estimator, we run the inference computation under different feature settings and use the inference time cost as the label value. While training the s-Estimator, we synchronize the model layer under different feature settings and use the communication time cost as the label value. Each estimator is trained with 330K samples, and the two estimators serve as the oracle to estimate the computation time (inference) and the communication time (synchronization) during the DP process (\S\ref{sec:dpp}).

\subsection{Dynamic Partition Planner (DPP)}
\label{sec:dpp}
To find the optimal partitioning scheme, DPP first constructs a search space and then runs a dynamic programming algorithm to search for the optimal partition scheme which can minimize the inference time.

\subsubsection{Search Space}
\begin{figure*}[!t]
	\centering
	\includegraphics[width=1\textwidth]{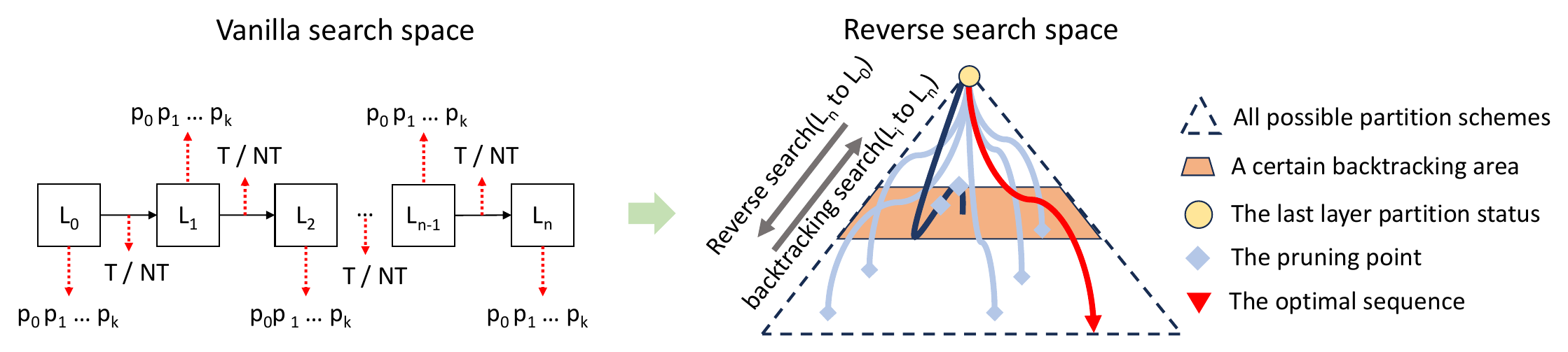}
	\caption{DPP's search space}
	\label{space}
\end{figure*}

The left side in Figure~\ref{space} sketches how DPP constructs the under-optimized search space (very large). Given a DNN with $n+1$ layers, denoted as $L_0$-$L_n$, the search starts from $L_n$. For each layer $L_i$, DPP makes the two-step choices, and $L_i$ is tagged with a pair $P_{i}=(p_i, t_i)$, $p_i\in\{\text{InH}, \text{InW}, \text{OutC}, \text{2D-grid}\}$ and $t_i\in \{\text{T}, \text{NT}\}$. DPP aims to find a sequence $S=[P_0, P_1,\cdots,P_n]$.

\Para{Step-1:} DPP needs to choose the dimension to partition $L_i$. We have $k$ different dimensions to consider (e.g. InH-based, InW-based, OutC-based, 2D-grid, etc) and DPP will choose one from them, and then continue to make choices in Step-2.

\Para{Step-2:} DPP still needs to decide whether or not the output of $L_i$ needs to be transmitted across edge nodes, in other words, DPP needs to choose between two modes. (1) Transmission (T) mode. The output of $L_i$ is not sufficient to serve as the input of $L_{i+1}$ because the input of $L_{i+1}$ still requires the boundary data (\S\ref{sec:trade-off}). As a result, between the computation of $L_{i}$ and $L_{i+1}$ will be a round of the inter-node communication for nodes to transmit the necessary data between each other. (2) Non-Transmission (NT) mode. The output of $L_i$ is sufficient to serve as the input of $L_{i+1}$ because $L_i$'s inference includes redundant computation. As discussed in \S\ref{sec:trade-off}, we need to consider this trade-off and flexibly decide whether or not to conduct redundant computation for each layer.

Since the typical DNN model can include tens or even hundreds of layers, simply enumerating every combination of partition schemes for each layer can lead to combinatorial explosion(explain in \S\ref{sec-Search Time Comparison}), we resort to the dynamic programming algorithm to search for the optimal partition scheme more efficiently.

\subsubsection{Dynamic Programming Algorithm}
\label{sec:optimal-eval}

Algorithm 1 describes the workflow of the dynamic program process in \sysname. As shown in the right side of Figure~\ref{space}, DPP starts the search from the last layer of the model. In general, DPP is trying to evaluate each optimal substructure during the traversal and finally find out the optimal structure (solution). Specifically, during reverse search (from $L_n$ to $L_0$), every time DPP considers a $P_i$ where $t_i=\text{T}$. At this time, the algorithm will backtrack (from $L_i$ to $L_n$) and generate a combined sequence, which contains multiple subsequences to be evaluated (as shown in Figure~\ref{fig:dpp-illustration}). For each subsequence, the starting point $P_{i}$ and the ending point $P_{j}$ will be T mode (i.e., $t_i=\text{T}$ and $t_j=\text{T}$). DPP will evaluate each subsequence one by one, take the optimal value and save it to the beginning $t_i$.

\Para{Key design-1: Reverse search}. Given the search space, DPP needs to decide the search order. A key point is that the sequence $S$ may contain multiple layers of consecutive NT cases (i.e., using redundant computation to save inter-node communication cost), and the impact of NT cases cascades and will be propagated forward through the layers ($L_i$ to $L_0$).
More specifically, if $P_{i+1}$ has $t_{i+1}=\text{NT}$, the inference of $L_i$ needs to perform redundant computation. Following that, if $P_i$ has $t_i=\text{NT}$, then inference of $L_{i-1}$ also needs to perform redundant computation, and the inference of $L_{i-1}$ needs to perform even more redundant computation.
In such cases, if the dynamic programming algorithm uses searches from $L_0$ to $L_n$, it will lead to a large number of redundant evaluations and cannot establish the optimal substructure. We realize this problem, and decide to adopt the reverse search order to run the dynamic programming algorithm: the search will be performed in order from $L_n$ to $L_0$ (line 19).  DPP first initializes the search space (lines 5-12) and assigns {T, NT} to $t_i$ of each layer $L_i$, $k$ groups per layer (lines 8-9). The last layer is different from previous layers because it must be transmitted after computation (lines 11-12). 

\begin{figure}[t]
	\centering
	\includegraphics[width=1\textwidth]{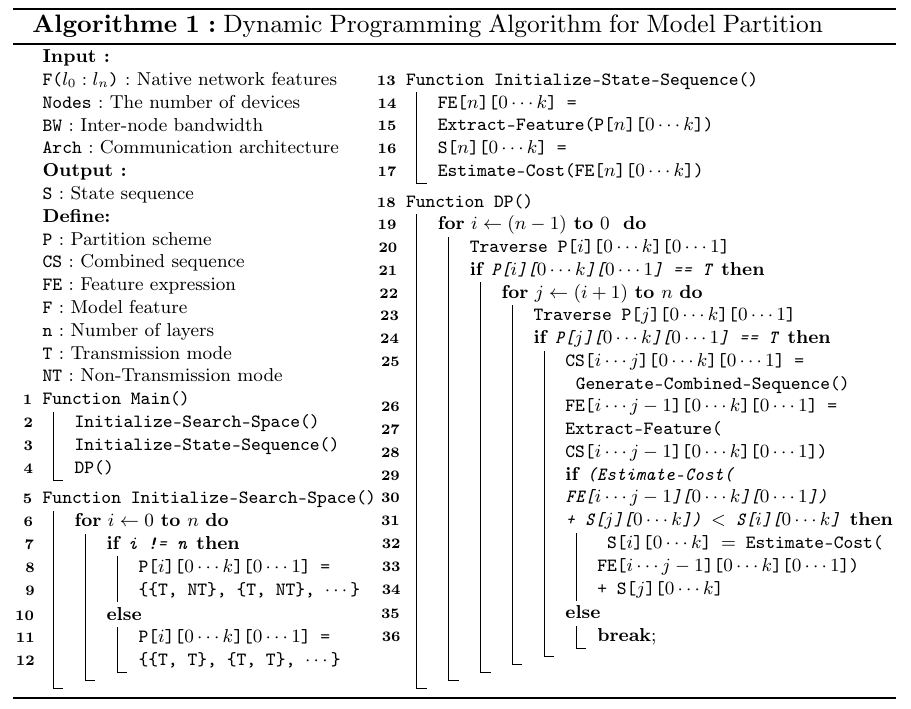}
	\label{algo:DPP-algo}
\end{figure}

\Para{Key design-2: Skip all NT states in reverse search}. During reverse search (i.e., $L_{n}\rightarrow L_{i}$), every time DPP only considers a $P_i$ where $t_i=\text{T}$ (line 21) and starts backtracking (line 22). This is because if DPP considers a $P_i$ where $t_i=\text{NT}$, the reverse search becomes a reverse full traversal search. In the process of backtracking, DPP does not evaluate the substructure ended with NT state (as shown in Figure~\ref{fig:dpp-illustration}) and only evaluates the substructure ended with T state (details explained next in Design 3).

\Para{Key design-3: Backtrack and generate combined sequences}. DPP will first consider $P_n$ where $t_n=\text{T}$. Then, DPP extracts its features and sends them to CE for estimation. Finally, DPP can get a total of $k$ costs of $P_n$ where $t_n=\text{T}$ and save to S[$n$][$0\cdots k$] (line 13-17). 

Then, during reverse search (from $L_n$ to $L_{n-1}$), DPP starts backtracking at $P_{n-1}$ (where $t_{n-1}=\text{T}$) and considers fully connecting $P_{n-1}$ (where $t_{n-1}=\text{T}$) and $P_n$ (where $t_n=\text{T}$). Then, DPP generates a combined sequence (a total of $k^2$ subsequences) and sends them to CE for estimation. DPP can obtain $k$ costs of $P_{n-1}$ (where $t_{n-1}=\text{T}$) and add them to the costs in S[$n$][$0\cdots k$] respectively. Based on the comparison between sequences, DPP finally obtains the final $k$ overall costs of $P_{n-1}$ (where $t_{n-1}=\text{T}$) and save to S[$n-1$][$0\cdots k$] (lines 19-36).

Next, DPP performs reverse search again (from $L_{n-1}$ to $L_{n-2}$) and starts backtracking at $P_{n-2}$ (where $t_{n-2}=\text{T}$). We find DPP can estimate the combination of $P_{n-2}$ (where $t_{n-2}=\text{T}$) and $P_{n-1}$ (where $t_{n-1}=\text{T}$), but can not estimate the combination of $P_{n-2}$ (where $t_{n-2}=\text{T}$) and $P_{n-1}$ (where $t_{n-1}=\text{NT}$). This is because we cannot estimate $P_{n-2}$ (where $t_{n-2}=\text{T}$). DPP chooses to save the information in $P_{n-1}$ (where $t_{n-1}=\text{NT}$) until DPP encounters the $P_n$ (where $t_n=\text{T}$). Then, DPP evaluates $P_{n-2}$ (where $t_{n-2}=\text{T}$) and $P_{n-1}$ (where $t_{n-1}=\text{NT}$), and then sums up the cost with $P_n$ (where $t_n=\text{T}$). Based on the comparison between other sequences, DPP finally obtains the final $k$ costs of $P_{n-2}$ (where $t_{n-2}=\text{T}$) and save to S[$n-2$][$0\cdots k$]. 

\Para{Why skip NT states?} In our case, DPP only evaluates the substructure $S_{sub}$ that starts with $P_{i}$ where $t_i=\text{T}$. Otherwise, if $S_{sub}$ that starts with $P_{i}$ where $t_i=\text{NT}$, the time cost of $S_{sub}$ is indeterminate because the computation workload of $L_{i}$ will vary depending on whether the previous layer $L_{i-1}$ chooses the Transmission mode or Non-transmission mode (i.e., whether $t_{i-1}=\text{T}$ or $t_{i-1}=\text{NT}$).\footnote{If $t_{i-1}=\text{NT}$, the inference of $L_{i}$ needs to conduct more computation to construct its input feature map (including the boundary data illustrated in \S\ref{sec:trade-off}). By comparison, if $t_{i-1}=\text{T}$, the inference of $L_{i}$  conducts less computation because the boundary data is obtained from the other edge nodes instead of redundant computation.} Therefore, DPP skips the sub-sequence starting with $P_i$ where $t_i=\text{NT}$. For every $S_{sub}$ that starts with $P_i$ where $t_i=\text{T}$, DPP records the current time cost of $S_{sub}$. The $S_{sub}$ will be used to compare with the other candidate sub-sequences and compose longer sub-sequences. Finally, the optimal structure (solution) will be established with a sequence $S$.

\Para{Piecing together}. In summary, DPP adopts multiple pruning strategies to reduce the search space and further improve the search efficiency. (1) During reverse search, DPP can improve search efficiency by ignoring evaluating $P_i$ where $t_i=\text{NT}$ (line 21). (2) DPP uses S[$j$][$0\cdots k$] to reduce the number of evaluations during the backtracking process (lines 32-34). (3) DPP incrementally generates the combination sequences and avoids much ineffective backtracking in the planning process (lines 22-36, through dynamic thresholds).

Based on our algorithm design, we have the following theorem. The proof is included in our technical report~\cite{flexpie-tech}.
\begin{theorem}[Optimality]
    Assuming Cost Estimator always reports the proper time cost for any given partition scheme, then DPP can output the optimal partition scheme for a given DNN model that yields the lowest time cost.
\end{theorem}

% \begin{proof}
% Let $P_{i,\cdots,j}$ denote a partition scheme for layers from $L_i$ to $L_j$. Let \textsc{estimate}($P_{i,\cdots,j}$) denote the time cost of the partition scheme reported by the estimator. The estimator only estimates $P_{i,\cdots,j}$ when $P_i.t_i=\text{T}$ and $P_j.t_j=\text{T}$, i.e., NT mode is skipped by setting such cases an $\inf$ cost.  Let $Cost(i, j)$ denote the optimal (lowest) time cost of $P_{i,\cdots,j}$. Then, based on our traversal algorithm (lines 19-36 in Algorithm 1), $Cost(i, j)$ satisfies the following recurrence.
% \begin{equation*}
%   Cost(i, j)=
%     \begin{cases}
%       \textsc{estimate}(P_{n,\cdots,n}) & \text{if } i=n \text{ and } j=n \\
%       \min\limits_{i\leq k<j}\{ \textsc{estimate}(P_{i,\cdots,k})+\textsc{estimate}(P_{k+1,\cdots,j}) \} & \text{otherwise}\\
%     \end{cases}       
% \end{equation*}
% DPP evaluates the recurrence of $Cost(i,j)$ in a bottom-up way. Note that since we evaluate $Cost(i, j)$ as the sequence length $j-i+1$ increases from 1 to $n$, all values for sub-problems referenced by the recurrence for $Cost(i, j)$ will have already been computed. At the end, DPP returns the optimal (lowest) value for $Cost(0, n)$, as well as its corresponding partition scheme, which in turn is the optimal partition scheme that yields the lowest time cost. 
% \end{proof}

%% file: experimental.tex
\section{Experimental Evaluation}
\label{sec-exp}
% \subsection{Experiment Setting}
\Para{Testbeds:} Our testbed includes 4 TMS320C6678\cite{tms} devices connected through SRIO. We have evaluated under different communication topologies, including Ring-based, PS-based and Mesh-based topologies.\footnote{Our evaluation shows similar performance results for Ring-based and Mesh-based topologies. Due to space limits, we omit the results of Mesh-based topology.} In addition, we run tests with various settings of inter-device bandwidth: 5Gbps, 1Gbps and 500Mbps.

\Para{Benchmarks:} 
We use four typical models as benchmarks to evaluate \sysname, namely MobileNet~\cite{howard2017mobilenets}, ResNet18~\cite{he2016deep}, ResNet101~\cite{he2016deep} and Bert~\cite{jawahar2019does}.

\Para{Baselines:} We compare \sysname with five baselines. Specifically, we compare \sysname with 3 fixed partition schemes, i.e., Xenos\cite{zhang2023xenos}(One-dim (OutC)), MoDNN\cite{mao2017modnn} and DeepSlicing\cite{zhang2021deepslicing}(One-dim (InH/InW)), DeepThings\cite{zhao2018deepthings}(2D-grid). Besides, we also consider two other baselines used by recent works. We refer to DINA~\cite{mohammed2020distributed} and PartialDI~\cite{dey2019embedded} and implement the layerwise partition, which uses different partition schemes for different model layers. We refer to AOFL~\cite{zhou2019adaptive} and EdgeCI~\cite{chen2024edgeci} and incorporate data fusion into the fixed model partition scheme.

\Para{Metrics:} We study the inference time cost and the DPP search time cost in our evaluation. We run each inference workload and each generated optimal sequence 1000 times and reported the average. Besides, in order to make an overall comparison between \sysname and the baselines across all the test cases, we define a summative metric called \emph{performance score}, to quantify the performance of each solution. \emph{performance score} is defined as follows. For a given model benchmark and a given testbed setting, we run both \sysname and the four baselines and obtain their inference time cost. Among the five time costs, denoted as $t_1-t_5$, we choose the smallest one $\min\{t_1,t_2,t_3,t_4,t_5\}$ as the reference, and the \emph{performance score} of each solution is computed as 
$Score_{i}=\frac{{\min\{t_1,t_2,t_3,t_4,t_5\}}}{t_i}$.

The value of each score will be a value between 0 and 1.0. If a partition strategy incurs a larger inference time, then its score will be smaller. The best partition strategy that yields the smallest inference time will be scored 1.0.

\subsection{Comparison on 4-node Testbed}
\label{4node}

\begin{figure*}[t]
	\centering
	\subfigure[MobileNet, ResNet18 on Ring and PS]{
		\label{MobileNet-ResNet18-4} 
		\includegraphics[width=0.99\textwidth]{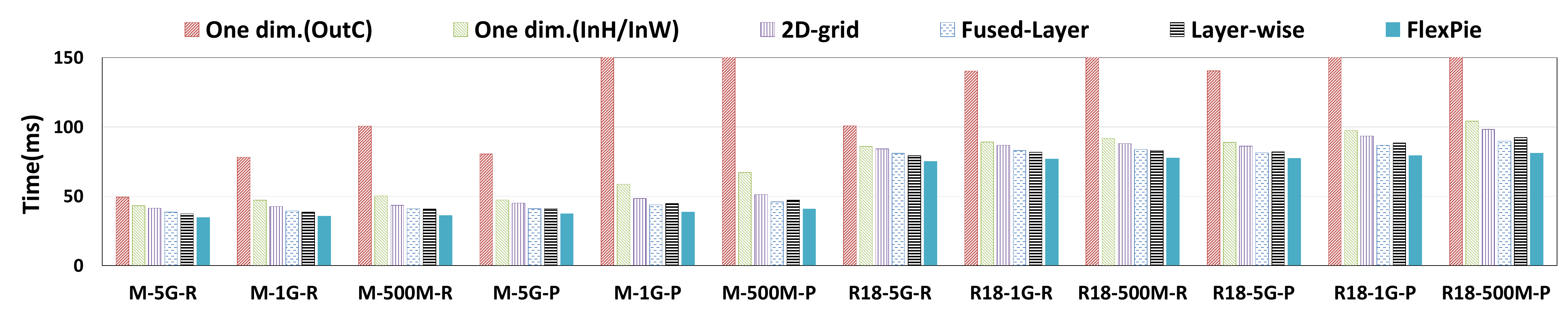}
	}
	\subfigure[ResNet101, Bert on Ring and PS]{
    	\label{ResNet101-Bert-4} 
    	\includegraphics[width=0.99\textwidth]{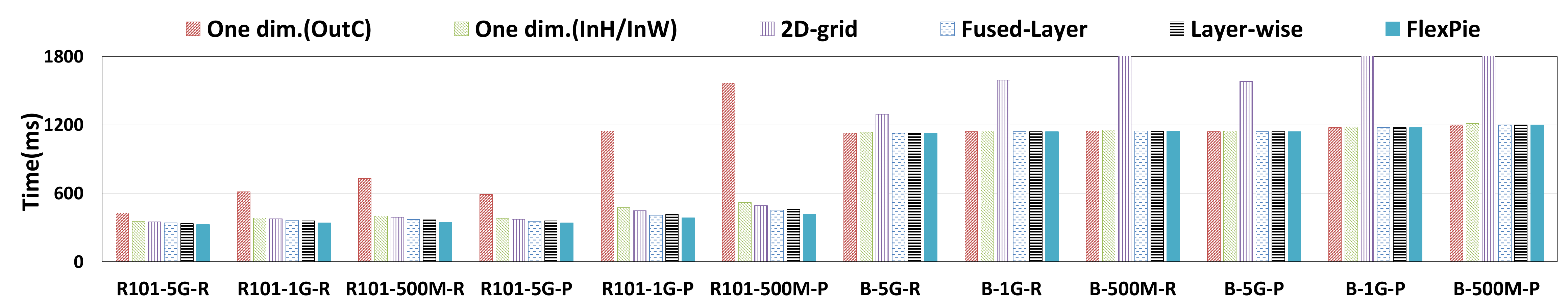}
	}
	\caption{Comparison on 4-node testbed}
	\label{Ring-4-node} 
\end{figure*}

We compare the inference time cost of different partition schemes on the 4-node testbed. Figure ~\ref{Ring-4-node} shows that, the 2D-grid partition performs best among the three baselines (OutC, InH/InW and 2D-grid) without layer-wise and fused-layer strategy. The OutC-based partition introduced non-trivial communication overheads. As illustrated in Figure~\ref{fig:model-partition}(c), with OutC-based partitioning, each node has to fetch the input feature maps from all the other three nodes to suffice its inference computation, so the performance is the lowest. The performance of one-dim.(InH/InW) is between 2D-grid partition and OutC-based partition. However, not all layers perform optimally with 2D-grid partition (as shown in Figure~\ref{fig:micro-bench}), because 2D-grid partitions will cause varying degrees of imbalanced calculation. For example, in MobileNet, when the input feature map is 14x14x512 or 7x7x512, the 2D-grid has unbalanced calculations because neither the InH dimension (14) nor the InW dimension (14) is a multiple of the number of nodes (4). The OutC-based partitioning may not be optimal for overall model partitioning, but it can achieve balanced computation (i.e., dividing the 512 channels evenly among the 4 nodes).

Compared with the One-dim and 2D-grid partition schemes, the layerwise partition introduces more flexibility and adopts different partition schemes for each layer. As a result, the layerwise partition yields better performance than the fixed One-dim and 2D-grid baselines. Different from layerwise partition, the layer fusion optimization accelerates the inference from the other direction. By fusing some model layers, the inference workflow can avoid redundant communication costs and in turn gain speedup. However, the previous works (e.g., DINA, EdgeCI) fail to incorporate both optimizations due to the explosive search space. Therefore, simply adopting layerwise optimization or fusion optimization cannot yield the best performance. By contrast \sysname can efficiently find the proper partitioning scheme through its dynamic optimization strategies (\S\ref{sec:dpp}). Therefore, \sysname distinctly reduces the inference time for a wide range of models under different typologies. 

\Para{Limitation:} \sysname provides little speedup when conducting inference with Bert. With a deeper dive, we realize that Bert model uses much matrix multiplication, but does not involve significant convolution computation as the other three benchmarks. Therefore, Bert already enjoys much easy parallelism in its nature. Neither layerwise flexibility nor inter-layer fusion brings distinct acceleration, leading to very close performance among different partition schemes.

\begin{figure*}[t]
	\centering
 	\subfigure[Score on 4node]{
		\label{Score-4} 
		\includegraphics[width=0.35\textwidth]{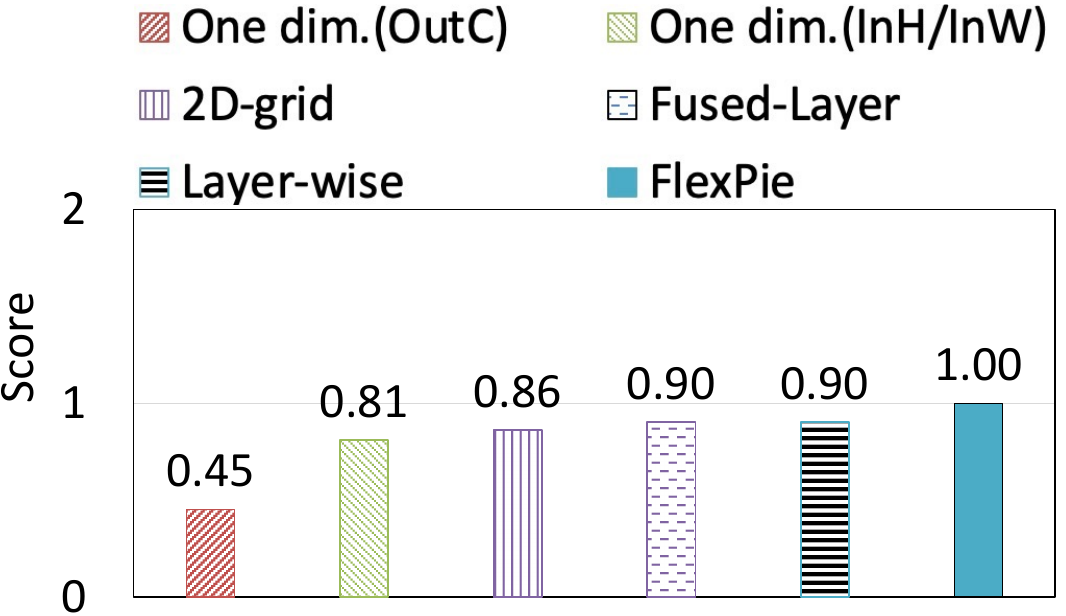}
	}
  	\subfigure[Score on 3node]{
		\label{Score-3} 
		\includegraphics[width=0.35\textwidth]{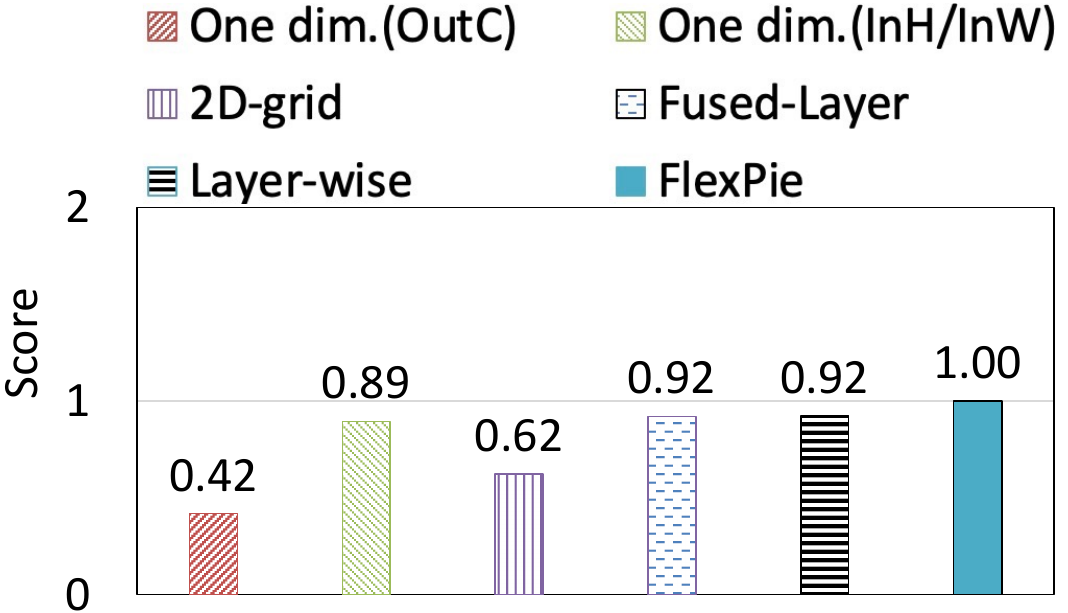}
	}
        \caption{Performance score comparison}
	\label{Score-4 and Score-4} 
\end{figure*}

\Para{Performance score:} \sysname achieves the highest performance score among all 5 solutions, and it outperforms the baselines with a speedup of 1.10-2.21$\times$.

\subsection{Comparison on 3-node Testbed}
\label{3node}

We then change the testbed setting and only use 3 nodes to run the inference tasks. In the 3-node test, the partition strategies including One-dim (InH/InW), One-dim (OutC) and 2D-grid still lead to imbalanced computation. But different from 4-node test,  the 2D-grid partition strategy becomes the worst case when running on the 3-node testbed. This is because, one node needs to undertake much more (i.e., twice as much) computation workload as the other two nodes.

Based on the comparative evaluation in both 3-node testbed and 4-node testbed, we also demonstrate that, there is no ``one-size-fits-all'' fixed partition scheme to provide general performance improvement. More specifically, One-dim (OutC) and 2D-grid partition may yield the best performance under different bandwidths or different architectures. Since such statistic partition schemes can only reach their ``sweet spots'' under specific settings, they hardly provide general performance benefits to distributed inference under varying testbed configurations. By contrast, DPP enables \sysname to adapt to different testbeds and automatically search for the appropriate partition strategies to yield high performance, making \sysname more desirable for practical deployment. 

\begin{figure*}[!t]
	\centering
	\subfigure[MobileNet, ResNet18 on Ring and PS]{
		\label{MobileNet-Ring-3} 
		\includegraphics[width=0.99\textwidth]{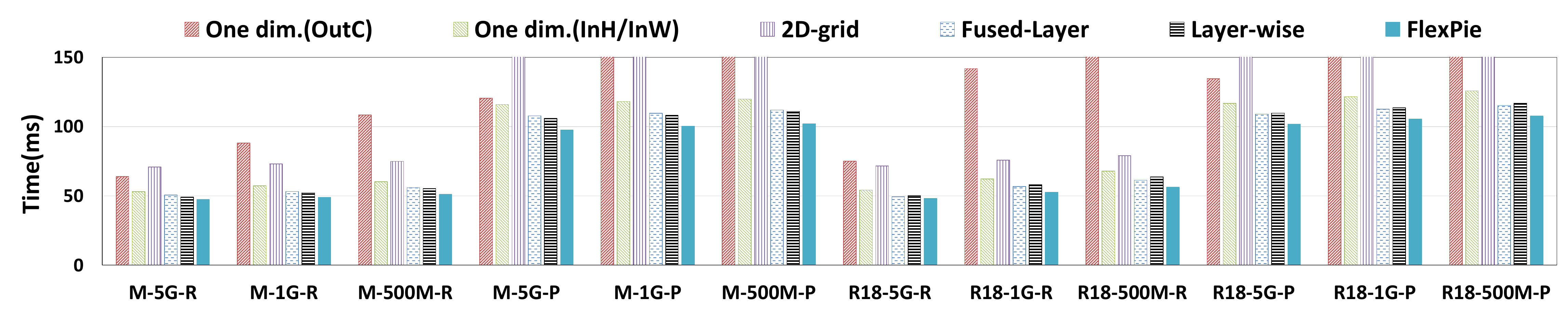}
	}
	\subfigure[ResNet101, Bert on Ring and PS]{
    	\label{ResNet18-Ring-3} 
    	\includegraphics[width=0.99\textwidth]{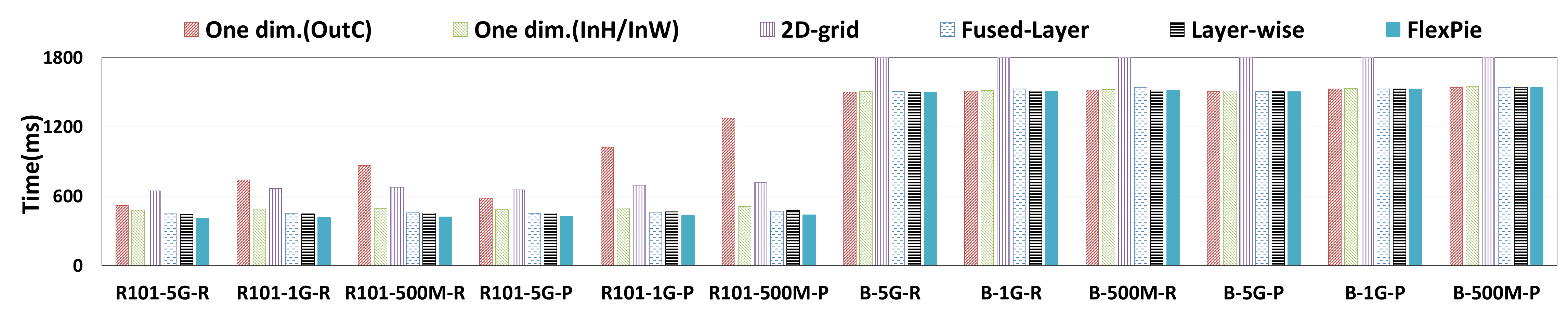}
	}
	\caption{Comparison on 3-node testbed}
	\label{Ring-3-node} 
\end{figure*}

\Para{Performance score:} \sysname achieves the highest performance score among all 5 solutions, and it outperforms the baselines with a speedup of 1.08-2.39$\times$.

%% file: related.tex
\section{Related Work}
\label{sec-related}
\Para{Distributed DNN Inference} Distributed DNN inference can be categorized into \emph{cloud-to-edge devices}~\cite{song2018situ,jeong2018ionn}, \emph{edge server-to-edge device}~\cite{mohammed2020distributed,shan2020collaborative}, \emph{cloud-to-edge server-to-edge device}~\cite{ren2021fine,xue2021eosdnn,lin2019distributed,dey2019offloaded} and \emph{edge device-to-edge device}~\cite{zhao2018deepthings,zhang2021deepslicing,mao2017modnn,zhou2019adaptive}. Among them, \emph{edge device-to-edge device} deploys the DNN on the local terminal device and performs DNN inference entirely in a local collaborative manner. This paradigm focuses on inference latency and energy consumption, but it can be applied to high mobility scenarios or some remote and harsh environments. In this paper, we focus on how to obtain lower inference latency through flexible combinatorial optimization in device-device collaborative inference scenarios.

\Para{Model Partitioning in Distributed Inference}
MoDNN~\cite{mao2017modnn} uses the one-dim to minimize the non-parallel data transmission time, but ignores the computational imbalance problem caused by the one-dim scheme. 
Deepslicing~\cite{zhang2021deepslicing} achieves a trade-off between computation and synchronization through the Proportional Synchronization Scheduler. But it also adopts one-dim scheme, which constrains its optimization opportunities. 
DeepThings\cite{zhao2018deepthings} uses a 2D-grid to perform model partitioning and implements a distributed work-stealing approach to enable dynamic workload distribution and balancing during inference runtime. However, it still suffers from the imbalanced distribution of computation tasks. In addition, many model partition strategies have been developed and applied in distributed training (e.g.,~\cite{vldb14-flexps,icde20-fela,elasticpipe}), and it will be an interesting future direction to study these partition strategies in the scenario of model inference. 

% \Para{Other Frameworks with Different Optimization Goals}
% CoEdge~\cite{zeng2020coedge} aims to minimize the energy consumption during edge-based inference. It tries to utilizes the available computation and communication resources at the edge and dynamically partitions the DNN inference workload according to the edge devices’ computing power and network conditions. 
% \cite{hadidi2018distributed} proposes a framework to save both energy consumption and memory cost. It leverages the computing power of multiple low-power robots to achieve efficient, dynamic, and real-time recognition. 
% \cite{goel2022efficient} designs a novel method that creates a parallel inference pipeline for computer vision problems that rely on hierarchical DNNs. This approach balances the load between cooperating devices and reduces communication costs to achieve higher throughput of video processing and minimize the inference latency.

%% file: conclusion.tex
\section{Conclusion}
\label{sec-conc}
This paper presents \sysname, a new inference work to employ multiple edge devices to jointly execute the model inference task. \sysname incorporates the flexible combinatorial optimization to automatically decide the partition scheme, which can reduce inference time cost, and yield desirable performance (up to 2.39$\times$) under various benchmark settings. 
% Our future work will consider (1) designing and implementing a more robust DDP algorithm which can adapt to the time-varying characteristics of the runtime environment (e.g., bandwidth fluctuation); (2) conducting a larger-scale comparative study of \sysname and other inference frameworks under more heterogeneous settings.

\section{Acknowledgment}
This work is supported by the National Key Research and Development Program of China (No. 2021ZD0110202).